\documentstyle[11pt,aaspp4]{article}
 
\begin{document}

\title{OH SATELLITE LINE MASERS AND AN AGN CANDIDATE IN M82}

\author{E. R. SEAQUIST and D. T. FRAYER}\affil{Astronomy
  Department, University of Toronto, Toronto, ON, M5S 3H8, Canada}
\author{AND} 
\author{D. A.  FRAIL}\affil{National Radio Astronomy Observatory,
  P.O. Box 0, Socorro, NM  87801}

\begin{abstract}
  We report the detection of OH satellite line masers at 1720 MHz and
  1612 MHz from the nuclear region of the starburst galaxy M82.  The
  observations were aimed at detecting 1720 MHz maser emission from the
  known radio emitting SNR in the nuclear region.  At 1720 MHz we detect
  six features above the 5$\sigma$ limit set by noise, four in emission
  and two in absorption.  Three of the emission features appear closely
  associated with known discrete continuum radio sources, and one of the
  absorption features is precisely coincident with the discrete
  continuum source 44.01+59.6. The latter source also exhibits strong
  1612 MHz emission at the same velocity. No other 1612 MHz features
  were detected. All of the 1720 MHz emission features are consistent
  with masers pumped by collisions with molecular hydrogen at densities
  between $10^3$~cm$^{-3}$ and 10$^5$~cm$^{-3}$, and T$_k$ between 50~K
  and 250~K.  The absorption and emission associated with the two
  satellite lines in 44.01+59.6, together with other evidence, points to
  the possibility that this source is the AGN in M82.
\end{abstract}


\section{INTRODUCTION}

M82 is the nearest (D$=3.2$ Mpc) and therefore most well studied
starburst galaxy, since its features may be examined at high luminosity
sensitivity and high linear resolution. The inner 1~kpc region contains
a large population of discrete luminous radio sources presumed to
represent primarily a population of radio emitting SNR (e.g., Kronberg,
Biermann, \& Schwab 1985; Muxlow et al. 1994).  Their high radio
luminosity stems from the interaction of the ejecta with the surrounding
dense molecular clouds responsible for star formation.  Here we present
a report of the first search for satellite OH maser line emission at
1720 MHz and 1612 MHz associated with this SNR population using the Very
Large Array.  Such features are observable in about 10\% of all galactic
SNR (Frail et al. 1996), and are also found to be associated with the
source Sgr~A East in the central region of the Galaxy (Yusef-Zadeh et
al.  1996).  These masers are probably associated with pumping by
collisions with H$_2$ molecules in SNR shocks, in accordance with the
models by Elitzur (1976).  The shock conditions present in the SNR of
M82 would therefore be expected to produce 1720 MHz masers similar to
those found in our Galaxy, and possibly with much higher luminosity.

\section{OBSERVATIONS}

Observations were made of M82 in the two OH satellite lines at 1720.527
MHz and 1612.231 MHz on November 17, 1996, using the Very Large Array of
the National Radio Astronomy Observatory.\footnote{The National Radio
  Astronomy Observatory is a facility of the National Science
  Foundation, operated under a cooperative agreement by Associated
  Universities, Inc.} The VLA correlator was configured to observe a
3.125 MHz bandwidth centered on a LSR velocity of 200~km~s$^{-1}$, close
to the systemic velocity of M82.  The bandwidth was divided into 128
spectral channels, providing a velocity resolution of 4.3~km s$^{-1}$and
4.5~km s$^{-1}$ at 1720 MHz and 1612 MHz, respectively.  Only right
handed circular polarization was recorded at each frequency.  The data
were amplitude, phase, and passband calibrated according to standard VLA
procedures using the AIPS software package.  Self--calibration of the
strong continuum of M82 allowed some additional improvement in the
dynamic range at both frequencies.  The continuum was subtracted from
the u--v visibility data using the AIPS task UVLIN which linearly
interpolates the continuum across the region containing the line
emission.  Maps were made using both natural and robust weighting, with
the latter giving an improvement in resolution of 40\% at the expense of
an increase of 20\% in the rms noise level.  Considerable interference
was detected at 1612 MHz from the Glonass satellite system.  After
careful editing, the resulting rms noise per channel is 0.7 mJy at 1720
MHz and 1.7 mJy at 1612 MHz.
 
The individual spectral channel maps were inspected to search for
features greater than five times the rms noise (i.e., 5$\sigma$), and
Gaussian fits were made to derive the parameters of these features. The
peak flux densities were determined from the naturally weighted data
while the positions were determined from the robust weighted data
(Table~1).  

\section{RESULTS}

The features which were found to be at the level of 5$\sigma$ or greater
at 1720 MHz are shown in Table 1 together with estimates of their
measured parameters.  Also shown are the flux densities (S), or their
upper limits, for any associated features at 1612 MHz.  None of the
features were significantly resolved by the beam ($1\farcs2 \times
1\farcs0$) of the robust weighted maps.  Figure~1(a) shows the locations
of these features on the 1720 MHz continuum map, which is also labeled
with the locations of the radio SNR cataloged by Kronberg et al. (1985).
Figure~1(b) shows the same features on a position--velocity (p--v)
contour plot of the CO emission along the major axis, adapted from Shen
\& Lo (1995).  The position of each feature is represented by its
location projected onto the major axis.  Figure 2 shows the individual
spectra from slices through the data cube at the maximum brightness of
the features on the channel maps.  Note that the data cube for the 6
features listed in Table~1 contains $3.4\times10^{5}$ independent
points, which leads to a Gaussian probability of 20\% for finding a
noise feature in the cube at the $5\sigma$ level or greater.  Thus, one
of the observed features could be spurious.

The 1720 MHz features (1), (2), (3), and (6) are seen in emission
whereas (4) and (5) are in absorption.  Features (1)--(4) appear
associated with discrete continuum sources, showing maximum
displacements of 1$\arcsec$ (15 pc projected separation).  Features (5)
and (6) do not appear associated with discrete continuum sources.
Instead, feature (5) appears in absorption against the background
continuum and the emission feature (6) is displaced well to the south of
the radio continuum disk and may be a noise feature, in accordance with
the probability computation above.  Note however that the latter does
appear superposed on a sub--mm dust continuum feature projecting
southward from the disk on the maps of Hughes, Gear, \& Robson (1994)
and Kruegel et al.  (1990).  Feature (2) is located roughly midway
between two SNR, approximately 1$\arcsec$ NE of the brightest and most
compact SNR 41.95+57.5, using the designation by Muxlow et al. (1994).
It is displaced in position by about 1$\arcsec$ and in velocity by about
$-7$~km~s$^{-1}$ with respect to the main line 1667 MHz maser m2
identified by Weliachew, Fomalont, \& Greisen (1984), and its flux
density at 1720 MHz is about is a factor of 8 or 9 weaker than feature
m2 at 1667 MHz.  Feature (4) is the most interesting one, since it is
unique in two respects. It exhibits both 1720 MHz absorption and bright
1612 MHz emission at the same velocity, and it is the only feature
precisely coincident with a discrete source, in this case 44.01+59.6.

Figure 1(b) shows that the velocities of the features follow the general
pattern associated with rotation of the molecular ring of radius 250 pc
in the nucleus of M82, although some of the them are displaced by up to
$\pm$100~km~s$^{-1}$, possibly signifying their association with
velocity disturbances in shocks.  Features (1)--(4) exhibit the same
kinematic signature in the p--v plot as the ridge of maximum brightness
in CO, signifying the association with star forming regions. Feature (2)
is located at the western end of a feature associated with the inner
nuclear ring of ionized gas, identified for example by Goetz et al. (1990)
and Achtermann \& Lacy (1995).

\subsection{The Physical Origin of the Features}

All of the 1720 MHz emission features are unresolved, and have T$_b >$
500~K, consistent with maser action. The 1612 MHz feature associated
with (4) has T$_b >$ 5000~K. The mean monochromatic luminosity
($=S\times D^2$) of the observed 1720 MHz masers ($3.9\times10^4$
Jy-kpc$^2$) is about two orders of magnitude greater than the most
luminous components of the counterparts in galactic SNR (e.g., $\sim
4\times 10^2$ Jy-kpc$^2$ for W28) observed by Frail et al. (1994) and a
factor of about 30 greater than the integrated luminosity of the masers
($\sim 1\times 10^{3}$ Jy-kpc$^2$) in the Galactic center observed by
Yusef--Zadeh et al. (1996).  This may possibly be accounted for by a
large number of components within the synthesized beam, which has a
projected diameter of about 17~pc at the distance of M82.

Features (1), (3), (5) and (6) have no associated main line masers.
Therefore they probably have the same origin as their counterparts
associated with galactic SNR, and are presumably pumped by collisions
with H$_2$ molecules.  Collisional pumping is most efficient when the
density and temperature are in the range and $10^{3} \rm{cm}^{-3} <
n(\rm{H}_{2}) < 10^{5} \rm{cm}^{-3}$ and 25~K $< T_{k} <$ 200~K
respectively (Elitzur 1976).  These conditions overlap readily with the
conditions found in molecular clouds in M82, though temperatures in the
higher part of this range would require shock heating, since the ambient
kinetic temperature for molecular clouds in M82 is about 45~K (e.g.,
Wild et al. 1992).  Feature (2) is associated with a main line maser,
and it is probably located in an HII complex in this region, since both
main line and satellite line masers are sometimes observed together in
galactic HII regions (Gaume \& Mutel 1987).  The absorption and emission
found together in feature (4), and their precise coincidence with one of
the discrete continuum source suggest a different origin.  We discuss
this feature separately.

\subsection{An AGN Candidate?}

The prominence of the brightest compact source 41.95+57.5 led to an
early suggestion that it may be a nuclear radio source associated with
the nucleus of M82 (Kronberg \& Wilkinson 1975).  However, its
structure, radio spectrum, and variability are consistent with its
current interpretation as a SNR (e.g., Huang et al. 1994).  Muxlow et
al. (1994) also suggested that the source 43.31+59.2, closer to the
2.2$\micron$ peak, might be associated with the AGN, but the absence of
a low frequency turnover, signifying a low free--free optical depth,
indicates that it is located on the near side of the ionized gas
distribution (Wills et al. 1997).  The latter authors also suggested
that the source 44.01+59.6 is a more likely AGN candidate.  It has a
strong turnover and a positive high frequency spectral index, expected
for an AGN source.  Our OH data support this latter suggestion, and we
now examine the properties of this source in more detail.
     
Both the 1720 MHz absorption and the 1612 MHz emission in feature (4)
are coincident in velocity and are spatially coincident with the source
44.01+59.6.  The continuum source is very compact with a diameter of
0.8~pc (Muxlow et al. 1994), and, as noted above, it has been suggested
as a possible AGN candidate by Wills et al. (1997) on the basis of its
unusual radio spectrum.  It is also one of four discrete radio sources
coincident with discrete X-ray sources identified by Watson, Stanger, \&
Griffiths (1984), and it is coincident with a super star cluster (e.g.,
O'Connell et al.  1994). It is near the 2.2 $\micron$ nucleus and the
kinematic center, and it is the site of the highest ratio of
HCO$^{+}$/CO in M82 (Seaquist et al., in preparation), signifying that
this region also has the highest molecular gas density and/or
ionization.  The former property is reminiscent of the high HCN/CO ratio
in the nucleus of NGC 1068 (Helfer \& Blitz 1995).  Finally, the
combined absorption/emission resembles the OH properties of the AGN in
Cen~A observed by van Langevelde et al.  (1995).  These authors
associate the OH features in Cen~A with a maser pumped by far IR in a
molecular cloud possibly in a circumnuclear torus associated the AGN.
Thus, our observations and other indicators lend support to the
suggestion by Wills et al. (1997) that 44.01+59.6 is the AGN of M82, and
that the OH features are associated with a circumnuclear disk or torus.

The opposite behavior in the two satellite transitions associated with
44.01+59.6 may be used to investigate the characteristics of the
absorbing cloud.  This effect occurs when the relevant hyperfine levels
in the ground state compete for the same IR pump photons associated with
transitions between the ground state and the first excited states in the
$^{2}\Pi_{3/2}~J=5/2$ (119$\micron$) and $^{2}\Pi_{1/2}~J=1/2$
(79$\micron$) states (e.g., Elitzur 1992).  When both rotational
transitions are optically thick, it leads to stimulated absorption at
1720 MHz and stimulated emission at 1612 MHz with equivalent optical
depths.  The required OH column density in the masering cloud is given
by $N(\rm{OH})/\Delta V > 10^{15}\,\rm{cm}^{-2}$(km~s$^{-1})^{-1}$,
where $\Delta V$ is the line width.  If the optical depth in the
masering lines is low, then amplification of the continuum would lead to
conjugate emission and absorption in the satellite lines, as is the case for
Cen~A.  The greater line/continuum ratio in emission in M82 indicates
$\tau > 1$, and the ratio of the line strengths then permits an estimate
of the optical depth and covering factor of the absorbing gas.  If we
assume conjugate optical depths against the discrete continuum source
at 1720 MHz and 1612 MHz, the inferred line/continuum ratios of $- 0.40
\pm 0.06$ and $2.69 \pm 0.14$ respectively (referred to the flux of the
discrete source) lead to an optical depth of 1.9 and a covering factor
of 0.5.
 
The foregoing considerations in turn lead to estimates of the absorbing
cloud properties as follows.  Using a line width of 13~km~s$^{-1}$
(Table 1), we obtain a lower limit to the OH column density of
$1.3\times 10^{16}$~cm$^{-2}$. Adopting an OH/H$_2$ abundance ratio of
$3\times 10^{-7}$, similar to that derived for Cen~A by van Langevelde
et al.  (1995), and a (spherical) cloud diameter of $d < 0.8$~pc, based
on the size of 44.01+59.6 and the covering factor of less than unity, we
obtain a lower limit on the H$_2$ density of $1.8\times 10^{4}$~cm$^{-3}$,
with an uncertainty of perhaps an order of magnitude.  This density is
similar to the densities in the molecular clouds in M82, and to the
absorbing cloud associated with OH absorption in Cen~A.

\section{SUMMARY}    

In conclusion, we find satellite line OH masers at 1720 MHz near three
SNR in M82, one maser south of the central disk, one feature in
absorption against the disk radio continuum, and a feature with both
absorption and emission at respectively 1720 MHz and 1612 MHz.  The
position of the feature showing both emission and absorption (feature
[4]) is precisely coincident with the discrete source 44.01+59.6.  We
conclude that it is probable that features (1), (2), and (3) are
produced by the collisional pumping by dense molecular gas associated
with star formation near the radio SNR, but are not necessarily a direct
consequence of the SNR themselves. The behavior in feature (4)
associated with 44.01+59.6 bears a strong similarity to that of the
satellite OH lines in the AGN of Cen A.  This similarity, along with
other evidence, supports the view that this discrete source is the AGN
in M82, and that the OH absorption features are produced by a
circumnuclear disk or torus.

We wish to thank Karen Wills for communicating the results of MERLIN
observations of M82 prior to publication.  This research was supported
by a grant to E.R.S. from the Natural Sciences and Engineering Research
Council of Canada.

\clearpage \figcaption[f1.ps]{(a) Grey scale representation of the 1720
  MHz continuum map showing the OH features listed in Table 1 (diamond
  symbols) and the discrete sources from Kronberg et al. (1985) (cross
  symbols).  Also marked is the 2.2 $\micron$ nucleus of the galaxy
  according to Dietz et al. (1986) (triangle symbol).  The image is
  rotated so that major axis of M82 is parallel to the horizontal axis.
  The position angle of the major axis was assumed to be $75\arcdeg$.
  (b) Features in Table 1 on a contour representation of the CO
  position-velocity map from Shen \& Lo (1995). Positions of the maser
  features shown are projected onto the major axis, and the positions
  are with reference to the 2.2 $\micron$ IR source at the nucleus.}

\figcaption[f2.ps]{The 1720 MHz spectra for each of the $5\sigma$
  features listed in Table 1.  In each case the spectra represent slices
  along the velocity axis of the spectral line cube at the position of
  peak brightness in the individual channel maps of the feature. The
  only 1612~MHz feature shown, and the only one detected, is represented by
  the dashed profile associated with feature (4).}

\clearpage

\begin{deluxetable}{crrrrrr}
\tablenum{1}
\tablewidth{0pt}
\tablecaption{Sources}
\tablehead{
\colhead{Feature} & 
\colhead{$\alpha$(B1950)} & 
\colhead{$\delta$(B1950)} & 
\colhead{1720 Line} & 
\colhead{1612 Line} &
\colhead{Velocity} &
\colhead{$\Delta V$ } \nl
&\colhead{($09^{\rm{h}} 51^{\rm{m}}$~$^{\rm{s}}$)}&
\colhead{($69\arcdeg 54\arcmin$~$\arcsec$)}&
\colhead{(mJy)}&
\colhead{(mJy)}&
\colhead{($\,\textstyle\rm{km~s}^{-1})$}&
\colhead{($\,\textstyle\rm{km~s}^{-1})$}
}
\startdata

1&40.80$\pm$0.09&59.4$\pm$0.3 
&$4.1\pm0.7$&$<5.1$&$132\pm4$&$<4.3$\nl 

2&42.09$\pm$0.07&58.2$\pm$0.2 
&$5.1\pm0.8$&$<5.1$&$85\pm4$&$17\pm4$\nl

3&43.33$\pm$0.07&59.4$\pm$0.2
&$3.1\pm0.9$&$<5.1$&$204\pm4$&$<4.3$\nl

4&44.02$\pm$0.02&59.5$\pm$0.1
&$-5.0\pm0.9$&$31.6\pm1.7$&$233\pm4$&$13\pm4$\nl

5&44.14$\pm$0.08&62.2$\pm$0.2
&$-3.6\pm0.7$&$<5.1$&$345\pm4$&$<4.3$\nl

6&44.49$\pm$0.12&44.8$\pm$0.3
&$3.5\pm0.8$&$<5.1$&$306\pm4$&$<4.3$\nl


\enddata
\end{deluxetable}

\newpage
\setcounter{figure}{0}
\begin{figure}
\includegraphics{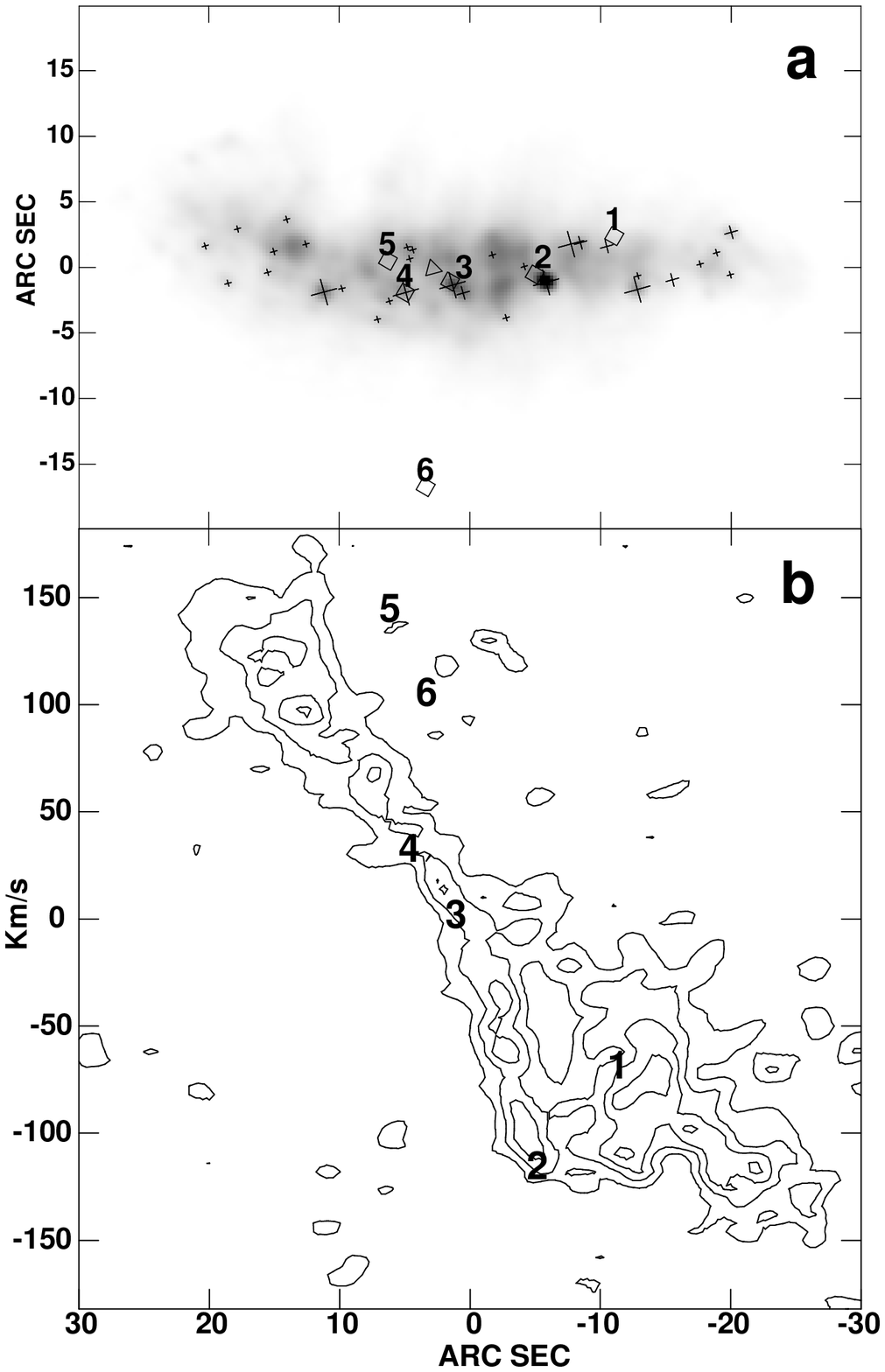}
\vspace*{8.in}
\caption{ }
\end{figure}

\begin{figure}
  \includegraphics{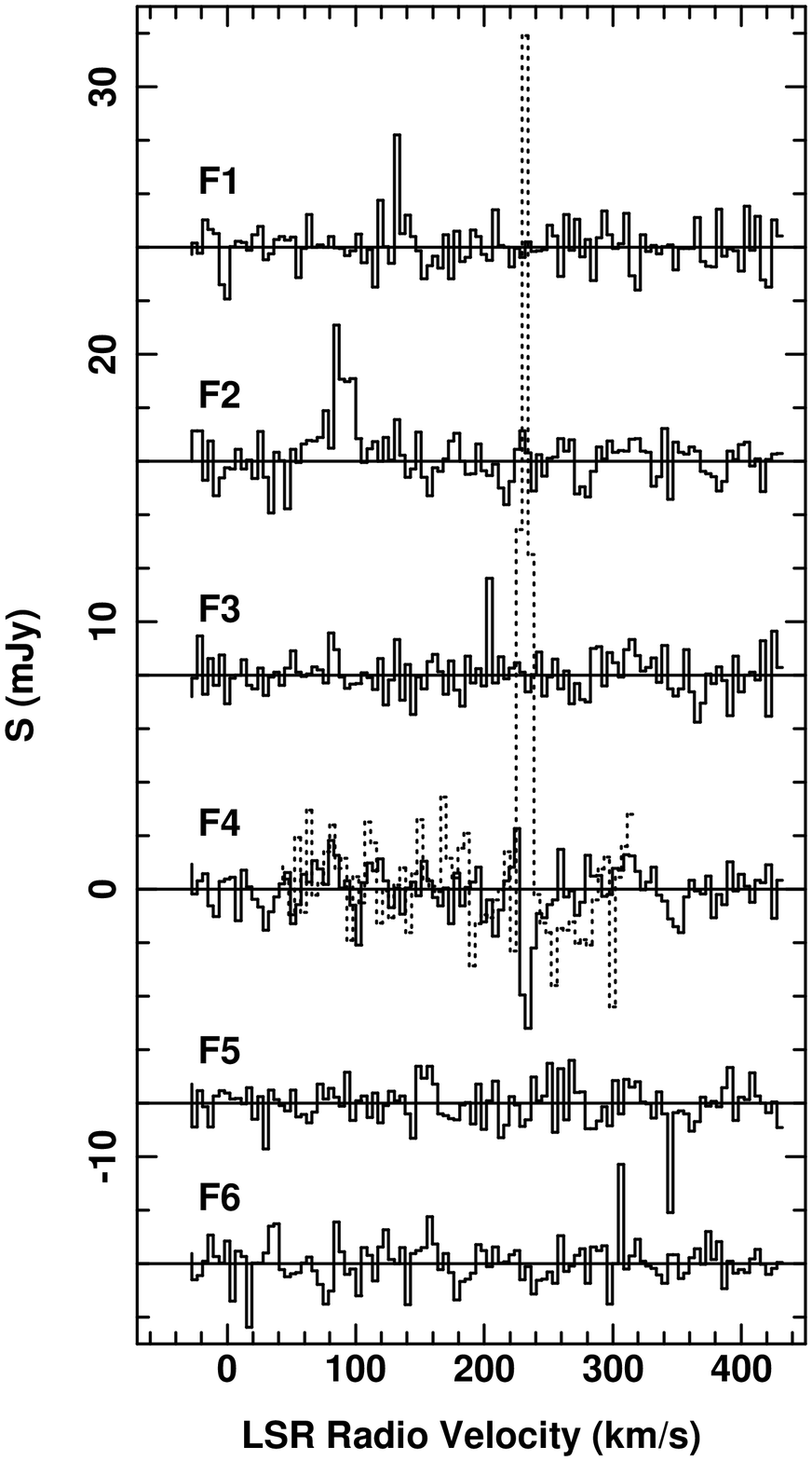} \vspace*{8in}
\caption{ }
\end{figure}


\begin{references}  


\reference{}Achtermann, J. M., \& Lacy, J. H. 1995, ApJ, 439, 163

\reference{}Dietz, R. D., Smith, J., Hackwell, J. A., Gehrz, R. D., \&
Grasdalen, G.  L. 1986, AJ, 91, 758

\reference{}Elitzur, M. 1976, ApJ, 203, 124

\reference{}Elitzur, M. 1992, Astronomical Masers (Dordrecht: Kluwer)

\reference{}Frail, D. A., Goss, W. M., Reynoso, E. M., Giacani, E. B.,
Green, A. J., \& Otrupeck, R.  1996, AJ, 111, 1651

\reference{}Frail, D. A., Goss, W. M., Slysh, V. I. 1994, ApJ, 424, L111
     
\reference{}Gaume, R. A., \& Mutel, R. L. 1987, ApJS, 65, 193

\reference{}Goetz, M., McKeith, C. D., Downes, D., \& Greve, A. 1990,
A\&A, 240, 52

\reference{}Helfer, T. T., \& Blitz, L.  1995, ApJ, 450, 90

\reference{}Huang, Z. P., Thuan, T. X., Chevalier, R. A., Condon, J. J.,
\& Yun, Q. F. 1994, ApJ, 424, 114

\reference{}Hughes, D. H., Gear, W. K., \& Robson, E. I.  1994, MNRAS, 270, 641

\reference{}Kronberg, P. P., Biermann, P., \& Schwab, F.  1985, ApJ, 291, 693

\reference{}Kronberg, P. P., \& Wilkinson, P. N. 1975, ApJ, 200, 430

\reference{}Kruegel, E., Chini, R., Klein, U., Lemke, R., Wielebinski,
R., \& Zylka, R.  1990, A\&A, 240, 232

\reference{}Muxlow, T. W. B., Pedlar, A., Wilkinson, P.M., Axon, D. J.,
Sanders, E. M., \& deBruyn, A. G.  1994, MNRAS, 266, 455

\reference{}O'Connell, R. W., Gallagher, G. S., Hunter, D. A., \& Colley,
W. N. 1995, ApJ, 446, L1

\reference{}Shen, J. \& Lo, K. Y.  1995, ApJ, L99

\reference{}van Langevelde, H. J., van Dishoeck, E. W., Sevenster, M., \&
Israel, F. P. 1995, ApJ, 448, L119

\reference{}Watson, M. G., Stanger, V., \& Griffiths, R. E. 1984, ApJ, 286, 144 

\reference{}Weliachew, L., Fomalont, E. B., \& Greisen, E. W.  1984, A\&A,
137, 335

\reference{}Wild, W., Harris, A. I., Eckart, A., Genzel, R., Graf, U.
U., Jackson, J. M., Russell, A. P. G., \& Stutzki, J. 1992, A\&A, 265, 447

\reference{}Wills, K. A., Pedlar, A., Muxlow, T. W. B., \& Wilkinson, P.
N. 1997, MNRAS, in press

\reference{}Yusef-Zadeh, F., Roberts, D. A., Goss, W. M., Frail, D. A.,
\& Green, A. J.  1996, ApJ, 466, L25

\end{references}
\end{document}